\documentstyle[aps,pre]{revtex}

\begin{document}
\title{Decay Theory of Double Giant Resonances$^{*}$}
\author{B.V.~Carlson$^{1,**}$, M.S.~Hussein$^{2,**}$ and
A.F.R.~de Toledo Piza$^2$}
\address{$^1$Departamento de F\'{\i}sica do Instituto Tecnol\'ogico da\\
Aeron\'autica - CTA\\
12228-900 S\~ao Jos\'e dos Campos, SP, Brazil}
\address{$^2$Instituto de F\'{\i}sica, Universidade de S\~ao Paulo,\\
C.P. 66318, S\~ao Paulo, 05315-970, Brazil}
\maketitle

\begin{abstract}
The decay theory of double giant resonances incorporating fluctuation
contributions of the Brink-Axel type is developed. The $\gamma $  and 
neutron emission decay of Double Giant Dipole Resonances
(DGDR) in $^{208}$Pb is discussed in connection with a recent measurement.
\end{abstract}


\begin{quote}
$*$~Supported in part by FAPESP.\\$**$Supported in part by CNPq.
\end{quote}


Recently, the decay properties of the double giant dipole resonance (DGDR)
in several nuclei has been investigated experimentally in Coulomb excitation
reactions at high energies \cite{emling0}. In particular, the neutron- and
$\gamma $-decay channels were looked at in Ref. \cite{emling}.
In analysing the data, the
authors rely on a model for the formation of the DGDR that involves the
sequential excitation of the two phonon state through the one phonon state.
Although it is concluded that the decay of the two phonons seem to follow
the harmonic model (namely the two phonons decay independently from each
other), we believe that a component essential to the analysis is missing,
since the integrated cross sections obtained from the decay data deviate
appreciably from the harmonic coupled channel calculations. The purpose of
this paper is to develop a new model for the decay of the DGDR using the
recently developed Direct~+~Fluctuation (DF) model of Ref. \cite{ccchp,ccchp2}
which reproduces the cross section value. We first give a brief description
of the DF model. It is argued in Ref. \cite{ccchp} that besides the,
direct, g.s. $\rightarrow$ one-phonon $\rightarrow$ two-phonon transition,
there is another contribution that arises from the coupling of the one-phonon
state to the complex background states followed by the excitation of a
Brink-Axel phonon.\cite{brax} The general structure of the two-step amplitude
is

\begin{equation}
T_{f0}=\langle\chi_f^{(-)},\varepsilon\mid V_{fi}\;({\rm d}+{\rm b})\;G_i\;%
{\rm d}\;V_{i0}\mid\chi_0^{(+)}\rangle \label{two-step}
\end{equation}

\noindent where $G_i$ is the propagator in the region of the GDR, ${\rm d}$
and ${\rm b}$ are projectors for the one phonon collective states and for
the background states responsible for the damping of the collective states
respectively. In writing Eq. (\ref{two-step}) it is being assumed that the
interaction $V_{i0}$, when acting on the initial state, will not excite the
background
states appreciably. These states will however be reached through the damping
mechanisms present in $G_i$, and will thus participate actively in the
second step. We next think of the propagator $G_i$ as split into a sum of
average parts which are essentially diagonal in the subspaces ${\rm b}$ and $%
{\rm d}$, $\bar{G}_b$ and $\bar{G}_d$ (with complex $Q$-value), plus an
additional fluctuation part with zero average. Thus

\begin{equation}
G_i=\bar{G}_b+\bar{G}_d+G_i^{{\rm fl}},\;\;\;\;\;\; \bar{G_i^{{\rm fl}}}=0
\label{gbarflu} \end{equation}

\noindent and the transition amplitude is similarly split as $T_{f0}= \bar{T}%
_{f0}+T_{f0}^{{\rm fl}}$ with

\begin{eqnarray}
\bar{T}_{f0}&=&\langle\chi_f^{(-)},\varepsilon\mid V_{fi}\;{\rm d}\; \bar{G}%
_d\;{\rm d}\;V_{i0}\mid\chi_0^{(+)}\rangle \equiv\bar{T}_{20},\label{tbar} \\
\nonumber \\
T_{f0}^{{\rm fl}}&=&\langle\chi_f^{(-)},\varepsilon\mid V_{fi}\;({\rm d} +%
{\rm b})\;G_i^{{\rm fl}}\;{\rm d}\;V_{i0}\mid\chi_0^{(+)}\rangle \equiv
T_{20}^{{\rm fl}}+T_{1^*0}^{{\rm fl}}\label{tfluc}
\end{eqnarray}

\noindent and the average cross section becomes

\begin{equation}
\bar{\sigma}=\mid \bar{T}_{f0}\mid ^2+\overline{\mid T_{f0}^{{\rm fl}}\mid ^2%
}
\end{equation}

\noindent with no cross term because of Eq. (\ref{gbarflu}).

The amplitude involving the averaged propagator is essentially the one given
by the coupled channels calculation described in Ref. \cite{bertu}. The
fluctuation part, Eq. (\ref{tfluc}), involves two different contributions. In
the first the final state is reached from the intermediate one-phonon state,
while in the second it is reached from the fine structure states to which
the intermediate one-phonon state decays. The importance of this
contribution stems from the possibility of collective dipole excitation of
these fine structure states. On general grounds it may be expected to be
proportional to the spreading width of the one ``cold''-phonon states, and
interference contributions with the two ``cold''-phonon amplitude to be
small. This leads to an excitation cross-section to the energy region of the
DGDR which is of the form

\begin{eqnarray}
\frac{d\bar{\sigma}^{(2)}}{d\varepsilon}&\simeq& \frac{1}{2\pi}\frac{%
\Gamma^{DGDR}}{(\varepsilon-E^{DGDR})^2+ (\Gamma^{DGDR})^2/4}\mid\bar{%
T}_{20}\mid^2+  \nonumber \\
&+&\frac{1}{2\pi}\frac{\Gamma^{GDR\downarrow}}{(\varepsilon-2E^{GDR})^2 +%
(\Gamma^{1^*})^2/4}\overline{\mid{T}_{1^*0}^{{\rm fl}}\mid^2}+ \frac{d%
\bar{\sigma}^{(2)\,{\rm fl}}}{d\varepsilon}  \nonumber
\end{eqnarray}

\noindent where the last term comes from $T_{20}^{{\rm fl}}$. This term is
quite small compared to the first two and we drop it in what follows. It is
known from work on giant dipole states in hot nuclei that the ``hot''-phonon
width $\Gamma ^{1^{*}}$ increases with excitation energy and then saturates.
However, at $E^{*}\equiv 2E^{GDR}$ it is essentially equal to $\Gamma ^{GDR}$%
. Moreover, for heavy nuclei $\Gamma ^{GDR\downarrow }\simeq \Gamma ^{GDR}$.
Denoting $\sigma _{dir} \equiv \left| \overline{T}_{20}\right| ^2$ and $%
\sigma _{f\ell } \equiv \left| \overline{T_{1*}^{\text{fl}}}\right| ^2$,
we have for the integrated cross section

\begin{equation}
\sigma ^{(2)}=\sigma _{dir}+\sigma _{fl}\label{sigdirflu}.
\end{equation}

When discussing neutron - or $\gamma $ - decay of the DGDR region, one has
to multiply each of the two terms on the RHS of Eq. (\ref{sigdirflu})
separately by the corresponding
branching ratio. For neutron emission, experience has shown that besides the
compound decay, there is also a preequilibrium component.\cite{bonetti,dias} 
Denoting
the branching ratio by $\rm{B}_{i\rightarrow n}^r=\frac{T_{i\rightarrow n}^r}{%
\stackrel{\sum T_{i\rightarrow f}^r}{f}}$ , with $T$ denoting the
appropriate transmission coefficient for the $n$- decay of state $i \left(
d^2\text{ or }d^{*}\right) $ through compound or preequilibrium
configuration $\left( r\right) $, we can write for neutron emission.

\begin{equation}
\sigma _n=\sigma _{dir}\left[ \rm{B}_{d^2\rightarrow n^{}}^{pre}\left(E_{d^2}
\right) +\rm{B}_{d^2\rightarrow n^{}}^{comp}\left( E_{d^2}\right)\right]
+\sigma _{fl}\left[ \rm{B}_{d^{*}\rightarrow n}^{pre}\left( E_{d^{*}}\right)
+\rm{B}_{d*\rightarrow n^{}}^{comp}\left( E_{d^{*}}\right) \right] .
\label{sign}
\end{equation}

We next turn to the $\gamma $ - decay. It has been established
\cite{beene} that the $\gamma $ - decay of the GDR is composed of a direct 
plus compound pieces. Thus

\begin{equation}
\rm{B}_{d\rightarrow \gamma }\text{ }\left( E_d\right) =
\rm{B}_{d\rightarrow \gamma}^{dir}\left( E_d\right) 
+\rm{B}_{d\rightarrow \gamma}^{comp}\left(E_d\right) .\label{bgdirflu}
\end{equation}

The decay of the DGDR proceeds in the following manner: the two dipole
phonons decay directly in a sequential manner; 
$\rm{B}_{d\rightarrow 2\gamma}^{dir} \left( E_{d^2}\right) 
=\left( \rm{B}_{d\rightarrow \gamma }^{dir}\left( E_d\right) \right) ^2$, 
one phonon decay directly, while the other
through the compound nuclear state at $E_d$; $\rm{B}_{d^2\rightarrow 2\gamma
}^{dir-comp} \left( E_{d^2}\right) =2\rm{B}_{d\rightarrow \gamma }^{dir} %
\left( E_d\right) \rm{B}_{d\rightarrow \gamma }^{comp} \left( E_d\right) $, and
finally the $d^2$ state simply mixes into the compound nucleus, which
subsequently decays; $\rm{B}_{d^2\rightarrow 2\gamma \text{ }}^{comp}\left(
E_{d^2}\right) $. Thus

\begin{equation}
\rm{B}_{d^2\rightarrow 2\gamma }\left( E_{d^2}\right) =
\left( \rm{B}_{d\rightarrow\gamma }^{dir}\text{ }\left( E_d\right) \right)^2
+2\rm{B}_{d\rightarrow \gamma}^{dir}\left( E_d\right) 
\rm{B}_{d\rightarrow \gamma }^{comp}\left( E_d\right)
+\rm{B}_{d^2\rightarrow 2\gamma }^{comp}\left( E_{d^2}\right) .
\end{equation}

To simplify the form of Eq. (9), 
it in tempting to assume that $\rm{B}_{d^2\rightarrow 2\gamma }^{comp} \left(
E_{d^2}\right) =\left( \rm{B}_{d\rightarrow \gamma }^{comp}\text{ }\left(
E_d\right) \right) ^2$, which would give for 
$\rm{B}_{d^2\rightarrow 2\gamma } \left( E_{d^2}\right) 
=\left( \rm{B}_{d\rightarrow \gamma }\left(E_d\right) \right) ^2$. 
However, for the moment, we will use Eq. (9) as it stands.

The next step is to obtain an expression for the $\gamma $-decay of the $%
d^{*}$ state. It is a very simple matter to convince oneself that a
collective Brink-Axel phonon first decays directly by emitting one $\gamma $,
followed by the compound decay from the $b$ space at $E_d$. There is also a
compound-compound component,

\begin{equation}
\rm{B}_{d^{*}\rightarrow 2\gamma }\left( E_{d^{*}}\right) 
=\rm{B}_{d^{*}\rightarrow\gamma }^{dir}\left( E_d\right) 
\rm{B}_{d\rightarrow \gamma }^{comp}\left(E_d\right) 
+\rm{B}_{d^{*}\rightarrow 2\gamma }^{comp}\left( E_{d^{*}}\right) 
\text{.}
\end{equation}

Collecting terms (Eqs. 9 and 10), we find for the $2\gamma $ - emission
cross section

\[
\sigma _{2\gamma }\left( E_{d^2}\right) =\sigma _{dir}\left[ \left(
\rm{B}_{d\rightarrow \gamma }^{dir}\left( E_d\right) \right) ^2+
2\rm{B}_{d\rightarrow \gamma }^{dir}\left( E_d\right) 
\rm{B}_{d\rightarrow \gamma }^{comp}\left(E_d\right) 
+\rm{B}_{d^2\rightarrow 2\gamma }^{comp}\left( E_{d^2}\right) \right] 
\]
\begin{equation}
+\sigma _{fl}\left[ \rm{B}_{d^{*}\rightarrow \gamma }^{dir}\left( E_d\right)
\rm{B}_{d\rightarrow \gamma }^{comp}\left( E_d\right) +\rm{B}_{d^{*}\rightarrow
2\gamma }^{comp}\left( E_{d^{*}}\right) \right] \label{sigam2}
\end{equation}

Equations (7) and (11) are the principal result of this work. In the following
we present a detailed calculation for the DGDR decay in $^{208}$Pb.

Calculations of the statistical decay of the giant resonance states were
performed using the code STAPRE.\cite{uhl} This code permits the calculation
of up to six sequential emissions from a compound nucleus, can calculate the
complete $\gamma$ cascade for each residual nucleus in the decay chain and
permits a simple pre-equilibrium decay calculation (no angular momentum or
isospin dependence).

The input data necessary for the calculations are the structural data of the
residual nuclei, such as discrete state characteristics and level density
parameters, as well as the transmission coefficients for particle emission
and partial widths for gamma emission. In the calculations of the decay of 
$^{208}$Pb described here, the level density parameters were estimated using
a shell-corrected fit to their systematic dependence on mass and charge
numbers. In particular, the level density parameters for $^{208}$Pb and 
$^{207}$Pb were taken to be $a_{208}=7.32$ MeV$^{-1}$ and 
$a_{207}=8.48$ MeV$^{-1}$. The discrete levels
used were taken from the ENSDF evaluated level files.\cite{ensdf} We
included the first 5 levels of $^{208}$Pb, up to E$_x$=3.71 MeV, and the
first 7 levels of the neutron-emission residue $^{207}$Pb, up to E$_x$=2.66
MeV. Neutron transmission coefficients were calculated using the global
optical potential parameters of Becchetti and Greenless.\cite{becgr} Charged
particle emission was neglected due to the high Coulomb barrier and
relatively low excitation energies involved. E1 gamma emission was
calculated using the Brink-Axel approximation with the the $^{208}$Pb
giant dipole resonance located at its observed position E$_d$%
=13.4 MeV with its observed width $\Gamma _d$= 4 MeV.\cite{emling} Higher
multipole gamma emission was calculated using Weisskopf single-particle
estimates.

To study the effects of pre-equilibrium emission, the initial particle-hole
configuration of each of the resonances was specified according to the
standard model of their nature. Thus, to study the decay of the single giant
dipole resonance, the initial excited nucleus population was taken to be in
a 1-particle-1-hole configuration with J$^\pi =1^{-}$ at an excitation
energy of E$_d$=13.4 MeV. Similarly, the double giant dipole resonance was
modeled as a 2-particle-2-hole configuration at E$_{d^2}$=26.8 MeV, with 1/6
of the population having J$^\pi =0^{+}$ and 5/6 having J$^\pi =2^{+}$. The
hot giant dipole was modeled as a 3-particle-3-hole configuration, in which
one of the particle-hole pairs is attributable to the giant dipole resonance
with the other two pairs being the lowest order contribution (in terms of
particle-hole pairs) of the the `melted ' giant dipole resonance which
furnishs the hot background. The excitation energy E$_{d^{*}}$ and spin
populations of the hot dipole resonance were taken to be the same as those
of the double giant dipole resonance.

The branching ratio obtained for single-step statistical decay of the single
giant dipole resonance to the ground state of $^{208}$Pb is 
$\rm{B}_{d\rightarrow \gamma}^{comp}=2.7\times 10^{-5}$. 
We found the branching ratio for occupation of
the $^{208}$Pb ground state after the entire $\gamma $ cascade to be only
about 10\% larger, 
$\rm{B}_{d\rightarrow \gamma -cascade}^{comp}=3.0\times 10^{-5}$. Thus, the
single-step Brink-Axel E1 emission dominates the statistical $\gamma $ decay
of the single giant dipole resonance. We note, however, that the direct 
$\gamma $ decay branch is much more important than the statistical decay
branch. The total branching ratio for $\gamma $ emission is actually several
orders of magnitude greater than the statistical ratio we have calculated, 
$\rm{B}_{d\rightarrow\gamma}=1.9\times 10^{-2}$.

We found the branching ratios for statistical decay of the double giant
dipole resonance d$^2$ and the hot dipole resonance d$^{*}$ to be very
similar. In both cases, the single-step decay to the ground state is
negligible, as would be expected given the initial spin populations (0$^{+}$
and 2$^{+}$) and the dominance of the Brink-Axel E1 decay mechanism. The
code STAPRE gives no direct information on double $\gamma $ emission, as
it furnishes only the single-step occupations and the occupations after the
complete cascade. However, we can infer a good deal of information from the 
cascade
occupations. For both the double and the hot giant resonances, the branching
ratio we obtained for occupation of the $^{208}$Pb ground state after the
statistical cascade is 
$\rm{B}_{d^2\rightarrow \gamma- cascade}^{comp}\approx  
\rm{B}_{d^*\rightarrow \gamma-cascade}^{comp}\approx 2.6\times 10^{-9}$.
This is about 3 times what we would estimate for the two-step emission
process, which would give a branching ratio of about 
$\left( \rm{B}_{d\rightarrow \gamma- cascade}^{comp}\right) ^2\approx
9.0\times 10^{-10}$. However, a closer look at the occupation probabilities
during the $\gamma $ cascade reveal that a good deal of the probability
passes through the J$^\pi =3^{-}$ state of $^{208}$Pb. This decay path
necessarily involves a final E3 $\gamma $ emission to reach the ground state
and is not the path of interest to us. If we look instead at the occupation
probability that decays directly from the continuum to the ground state, we
get a better estimate of the double $\gamma $ emission. For this
(approximately 2$\gamma$)
partial $\gamma $ cascade, we indeed find excellent agreement with our estimate
based on the single giant resonance decay,
\begin{equation}
\rm{B}_{d^2\rightarrow 2\gamma}^{comp}
\approx \rm{B}_{d^*\rightarrow 2\gamma}^{comp}\approx 
(\rm{B}_{d\rightarrow \gamma}^{comp})^2\approx 1.0\times 10^{-9}\,.
\end{equation}

Using this relation, we can reduce the expression for the 
$2\gamma$-emission cross section,
Eq. (\ref{sigam2}), to 
\begin{equation}
\sigma _{2\gamma }\left( E_{d^2}\right) \approx \sigma _{dir}\left(
\rm{B}_{d\rightarrow \gamma }\left( E_d\right) \right) ^2
+\sigma _{fl}\,\rm{B}_{d^{*}\rightarrow \gamma }\left( E_d\right)
\rm{B}_{d\rightarrow \gamma }^{comp}\left( E_d\right) \,
\end{equation}
where we have made use of Eq. (\ref{bgdirflu}). We now use the fact the the
branching ratio for compound $\gamma$ emission is much smaller than that for
direct emission, 
$\rm{B}_{d\rightarrow \gamma}^{dir} >>\rm{B}_{d\rightarrow \gamma}^{comp}$,
to neglect the fluctuation term, which yields,
\begin{equation}
\sigma _{2\gamma }\left( E_{d^2}\right) \approx \sigma _{dir}\left(
\rm{B}_{d\rightarrow \gamma }\left( E_d\right) \right) ^2\,.\label{sigamp}
\end{equation}
As the direct double excitation cross section $\sigma_{dir}$ is what has been
called the harmonic term, we conclude that the $2\gamma$-emission cross 
section should approximately agree with its harmonic estimate. We observe
that this is consistent with the experimental value of Ref. \cite{emling}, 
$\sigma _{2\gamma }/ \sigma _{2\gamma }^{harm}=1.25(40)$,
which within the experimental error, is equal to our value of 1. We emphasize, 
however, that the $2\gamma$-emission cross section is not
proportional to the summed double and hot dipole resonance excitation cross
section. As is clear from the discussion leading to 
Eq. (\ref{sigamp}), the $2\gamma$-emission from the fluctuating term is greatly
suppressed relative to that from the direct term. We thus expect the 
$2\gamma$-emission cross section to be proportional to the direct double
excitation cross section, $\sigma_{dir}$, alone.

Turning now to the neutron emission, we first observe that
the branching ratio for pre-equilibrium emission from the initial particle-hole
configuration can be interpreted as the branching ratio for ``direct'' emission
(as contrasted with statistical emission) from the giant resonance.
Following this reasoning, we find that, of the 4.4\% pre-equilibrium
fraction of the neutron emission from the $^{208}$Pb single giant dipole
resonance, 65\% is emitted from the 1-particle-1-hole configuration. This
yields a branching ratio of 
$\rm{B}_{1p-1h}^{d\rightarrow n}\approx 2.6\times 10^{-2}$ for
``direct'' neutron emission, which is of the same size as the branching ratio
for direct $\gamma $ emission.  However, it is extremely small compared to
the branching ratio of the dominant statistical neutron emission.

Pre-equilibrium neutron emission is substantially more important in the
decay of the double and the hot giant resonances. We find that 20\% of the
neutron emission from the double giant dipole resonance is pre-equilibrium
emission. For the hot giant dipole resonance, 15\% of the neutron emission
is pre-equilibrium emission. We can estimate the branching ratio for direct
neutron emission from the double giant resonance by observing that, of the
20\% pre-equilibrium fraction of the neutron emission, 30\% is emitted from
the 2-particle-2-hole state. We thus obtain a branching ratio for ``direct''
neutron emission of $\rm{B}_{d^2\rightarrow n}^{dir}\approx 6.0\times 10^{-2}$.
We observe
that this value of 6\% has the pleasing property of being close to the
difference between the pre-equilibrium fractions for emission from the
double and the hot giant dipole resonances. This lends force to its
interpretation as the ``direct'' component of the neutron emission. For the
case
of the double giant dipole resonance, the remaining pre-equilibrium emission
fraction of about 15\% corresponds to the emission that passes through the
hot dipole resonance, that is, through the configurations in which at least
one of the coherent dipole phonons has ``melted'' into the background.

Independent of the division of neutron emission into pre-equilibrium and
equilibrium components, the total branching ratio for neutron emission is
approximately the same for both the double and the hot giant dipole resonances,
$\rm{B}_{d^2\rightarrow n}\approx \rm{B}_{d^*\rightarrow n}$. We can then write
the neutron emission cross section, Eq. (\ref{sign}), as
\begin{equation}
\sigma _n \approx (\sigma _{dir} + \sigma _{fl})\rm{B}_{d^2\rightarrow n}\,.
\end{equation}
The neutron emission cross section thus reflects the summed double and 
hot giant dipole excitation cross section. This result is also consistent
with the experimental value of Ref. \cite{emling}, 
$\sigma _{n}/ \sigma _{n}^{harm}=1.06(35)$, which, within the experimental
error, is equal to our estimate,  
$\sigma _{n}/ \sigma _{n}^{harm}= (\sigma_{dir}+\sigma_{fl})/\sigma_{dir}
\approx 1.33$, where the numerical value of the cross section ratio was
taken from Ref. \cite{emling}.

We conclude from our analysis that the $\gamma$ and neutron emission decay
modes can make an observable distinction between the double and hot giant
dipole resonances. In $^{208}$Pb, the $2\gamma$ emission comes predominantly
from the double dipole resonance, while neutron emission comes about equally
from the double and hot dipole resonances. We also find that pre-equilibrium
neutron emission from the double dipole resonance is slightly enhanced
over that from the hot dipole resonance and, in both cases, is greatly enhanced
when compared to the pre-equilibrium emission from the single giant dipole
resonance.

\end{document}